\documentclass[aps,prl,twocolumn,superscriptaddress,nofootinbib]{revtex4-2}

\usepackage{amsmath}
\usepackage{graphicx}
\usepackage{bm}
\usepackage{xcolor}

\begin{document}

\title{Roton Instability in Quantum Droplets with Finite-Range Soft-Core Interaction}

\author{Avra Banerjee\thanks{Emails: avrabanerjee1@gmail.com, avra.pcs@bose.res.in}}

\affiliation{Department of Physics of Complex Systems, 
S. N. Bose National Centre for Basic Sciences, 
Block JD, Sector III, Salt Lake, Kolkata 700106, India}
\footnotetext{Emails: avrabanerjee1@gmail.com, avra.pcs@bose.res.in}

\begin{abstract}

We investigate the emergence of roton instability in self-bound quantum droplets interacting via a finite-range soft-core potential modeled by a Heaviside step interaction. The ground-state properties are obtained by solving the extended Gross–Pitaevskii equation including Lee–Huang–Yang corrections with a nonlocal interaction term. The collective excitation spectrum reveals the formation and progressive softening of a roton minimum as the interaction strength and range increase. When the roton energy approaches zero, the system becomes unstable, signaling the onset of density modulation. The rotonic behavior is further characterized through the static structure factor, which exhibits pronounced peaks at the roton momentum.

\end{abstract}

\maketitle

\section{Introduction} Ultracold atomic gases provide a versatile platform for studying strongly correlated quantum many-body phenomena under highly controllable conditions. A major recent breakthrough in this field is the theoretical prediction of self-bound quantum droplets stabilized by quantum fluctuations~\cite{petrov2015} and their subsequent experimental observation in ultracold Bose mixtures and dipolar gases~\cite{barbut2016,schmitt2016,cabrera2018,derrico2019,semeghini2018}. These droplets represent a remarkable state of matter where mean-field attraction is balanced by repulsive beyond-mean-field corrections originating from Lee–Huang–Yang (LHY) quantum fluctuations~\cite{petrov2015,bottcher2021,blakie2020}. More comprehensive discussions on quantum droplets can be found in Refs.~\cite{rev1,rev2,rev3,rev4}. Quantum droplets were first realized experimentally in mixtures of weakly interacting Bose gases with attractive interspecies interactions~\cite{cabrera2018,semeghini2018}. Similar phenomena were later observed in strongly dipolar condensates of magnetic atoms such as dysprosium and erbium~\cite{kadau2016}. Dipolar systems exhibit long-range and anisotropic interactions that fundamentally modify the excitation spectrum of the condensate~\cite{lahaye2009,baranov2012,yukalov2018}. In particular, they can support roton–maxon dispersion relations analogous to those originally introduced in superfluid helium. The concept of a roton excitation plays a central role in the physics of strongly interacting quantum fluids. Originally proposed to explain the excitation spectrum of superfluid helium, the roton minimum is now generally interpreted as a precursor to density ordering or crystallization~\cite{nozieres2004}. In ultracold atomic systems, roton-like excitations can emerge when interactions possess significant momentum dependence. For instance, dipolar Bose–Einstein condensates exhibit roton minima due to the competition between short-range contact interactions and long-range dipole–dipole forces~\cite{santos2003,odell2003,ronen2007}. Experimental signatures of roton softening have been observed in dipolar gases~\cite{chomaz2018,petter2019,schmidt2021}. The softening of a roton mode signals an instability toward spatial density modulation and can lead to the formation of new quantum phases. In dipolar gases, this mechanism is closely related to the emergence of droplet arrays and supersolid states~\cite{kadau2016,tanzi2019,bottcher2019,chomaz2019,guo2019}. Supersolids combine frictionless superfluid flow with spontaneous breaking of translational symmetry~\cite{andreev1969,leggett1970}. Recent experiments have demonstrated supersolid behavior in dipolar condensates~\cite{tanzi2019,chomaz2019,norcia2021,sohmen2021,roccuzzo2020,gallemi2020, tengstrand2021}. Roton excitations are not restricted to dipolar interactions and can also arise in systems with finite-range soft-core potentials. Such interactions can be engineered experimentally using Rydberg dressing~\cite{henkel2010,hsueh2013,zhang2022}. The resulting effective interaction exhibits a plateau at short distances and decays at larger separations, producing a soft-core potential that strongly modifies the collective excitation spectrum~\cite{cinti2014,cinti2014b,rossotti2017}. In theoretical studies, simplified model potentials are often employed to isolate the essential effects of interaction range. Among these, the Heaviside step potential provides a convenient representation of a finite-range soft-core interaction~\cite{macri2013,boninsegni2012,saccani2011,rakic2024}. Despite its simplicity, this model captures the essential physics associated with momentum-dependent interactions and roton formation~\cite{rakic2024}. In the present work, we investigate the emergence of roton instability in quantum droplets interacting through a finite-range soft-core potential modeled by a Heaviside step interaction. The system is described within the framework of the extended Gross–Pitaevskii equation incorporating Lee–Huang–Yang corrections following Petrov’s formulation~\cite{petrov2015}. The ground state is obtained numerically, and the excitation spectrum is analyzed to identify roton modes and their relation to density correlations in the droplet.
\begin{figure*}[t]

\includegraphics[width=0.42\textwidth]{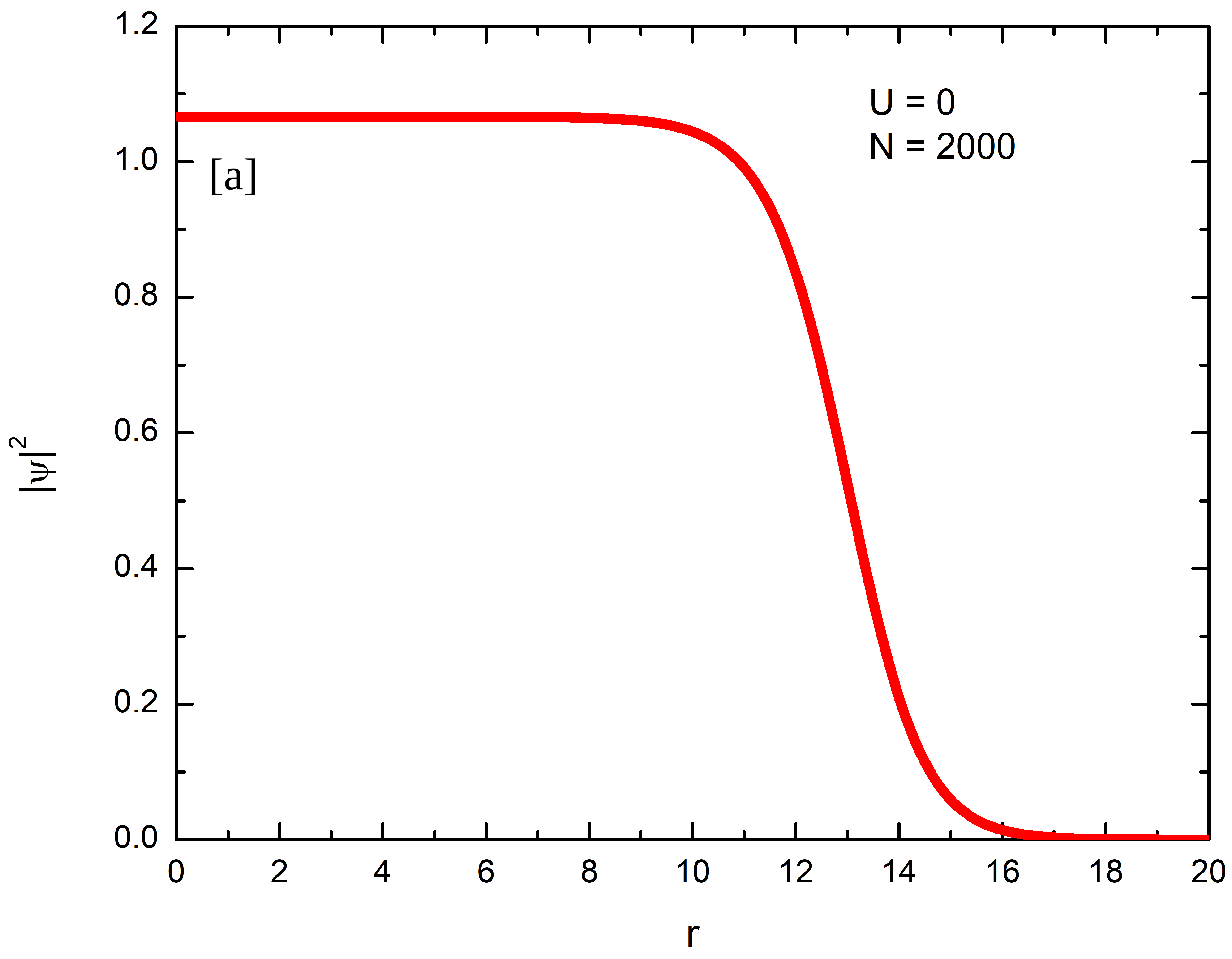}
\includegraphics[width=0.42\textwidth]{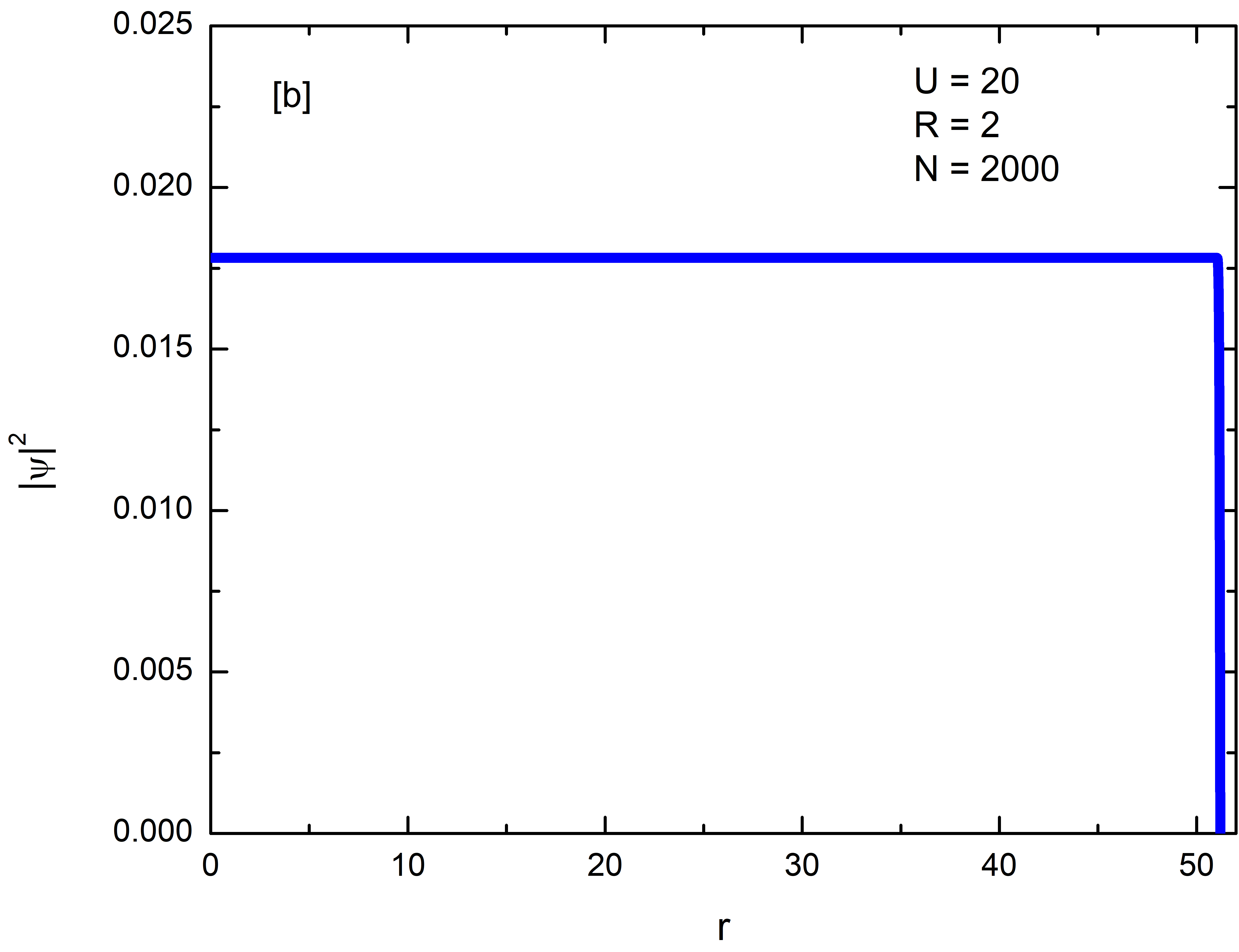}

\includegraphics[width=0.42\textwidth]{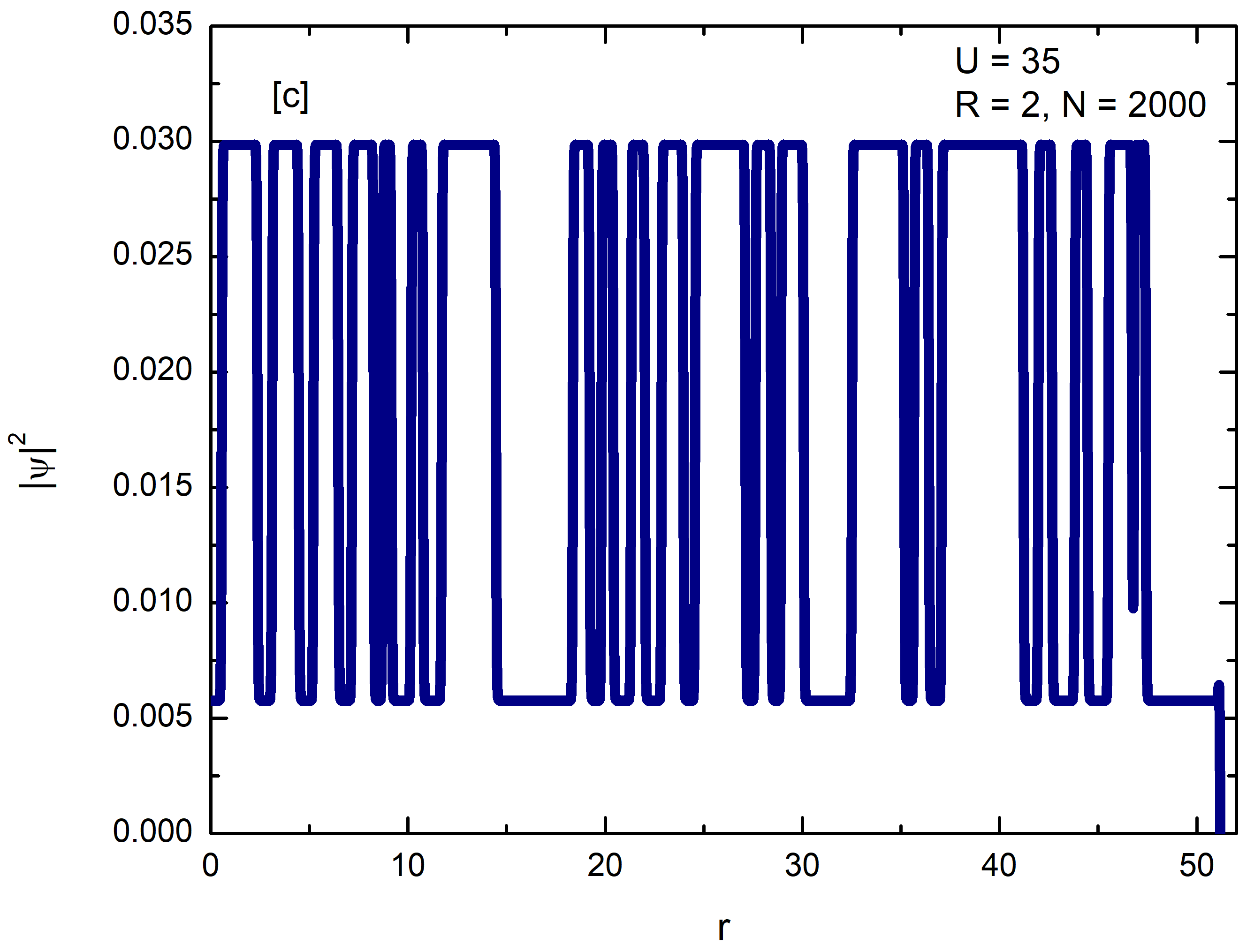}
\includegraphics[width=0.42\textwidth]{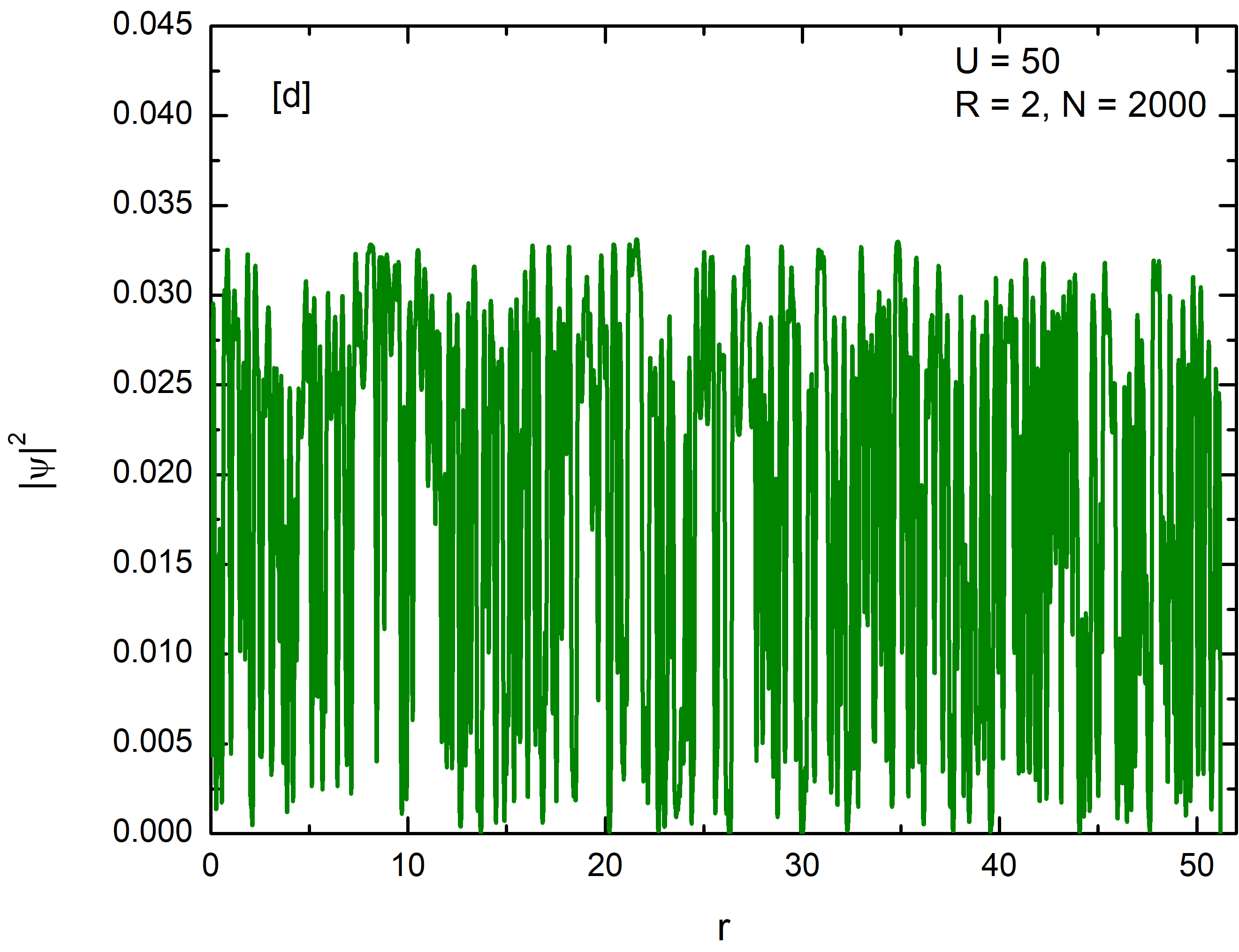}

\caption{
Density profile $|\psi|^2$ of a self-bound quantum droplet (obtained by solving equation (\ref{single_gp})) under the influence of a nonlocal Heaviside-type interaction with range $R=2$ and total particle number $N=2000$ for different interaction strengths $U$. (a) In the absence of nonlocal interaction, the droplet exhibits a characteristic flat-top density profile with a well-defined boundary. (b) For $U=20$, the nonlocal interaction spreads the droplet over a larger spatial region while the density remains nearly uniform.   The density decreases rapidly but continuously near the droplet surface, giving the appearance of an abrupt edge in the plotted scale.   (c) For $U=35$, these modulations become stronger and extend across the droplet, indicating significant perturbation of the uniform density background. (d) For $U=50$, the density distribution becomes highly irregular and random, suggesting fragmentation of the droplet structure under strong nonlocal interaction.}
\label{fig:droplet_density}
\end{figure*}
\section{Model and calculations}

 We consider a self-bound binary quantum droplet described by the effective single Gross--Pitaevskii equation, with contact interactions supplemented by an additional finite-range soft-core interaction,  modeled by the step potential. Such interactions remain approximately constant within a finite radius and vanish beyond it. A convenient representation
of the step potential is \cite{macri2013,boninsegni2012,saccani2011,rakic2024}

\begin{equation}
V(r)=V_0\,\Theta(R_c-r),
\end{equation}

\begin{figure*}[t]
\centering
\includegraphics[width=0.45\textwidth]{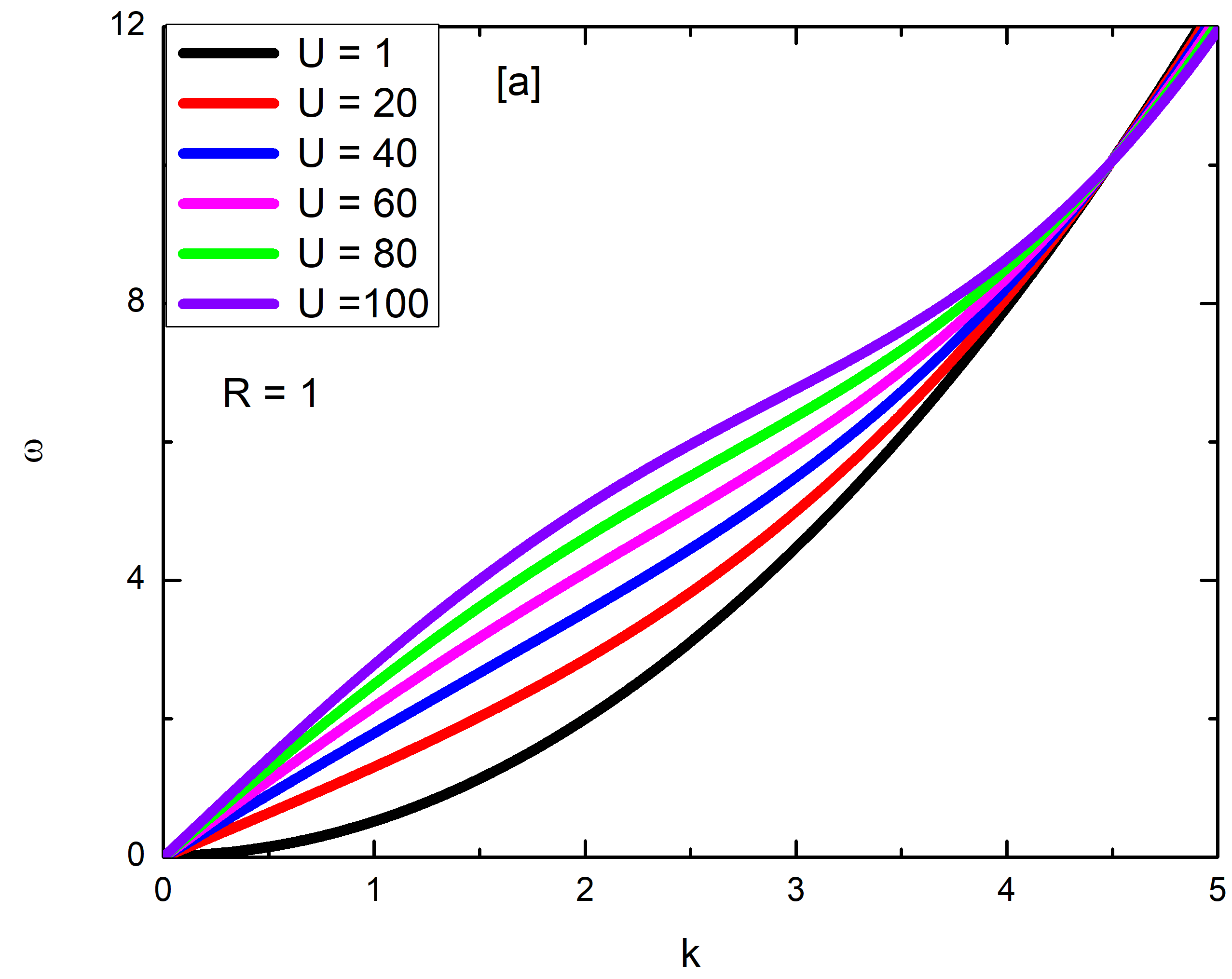}
\includegraphics[width=0.45\textwidth]{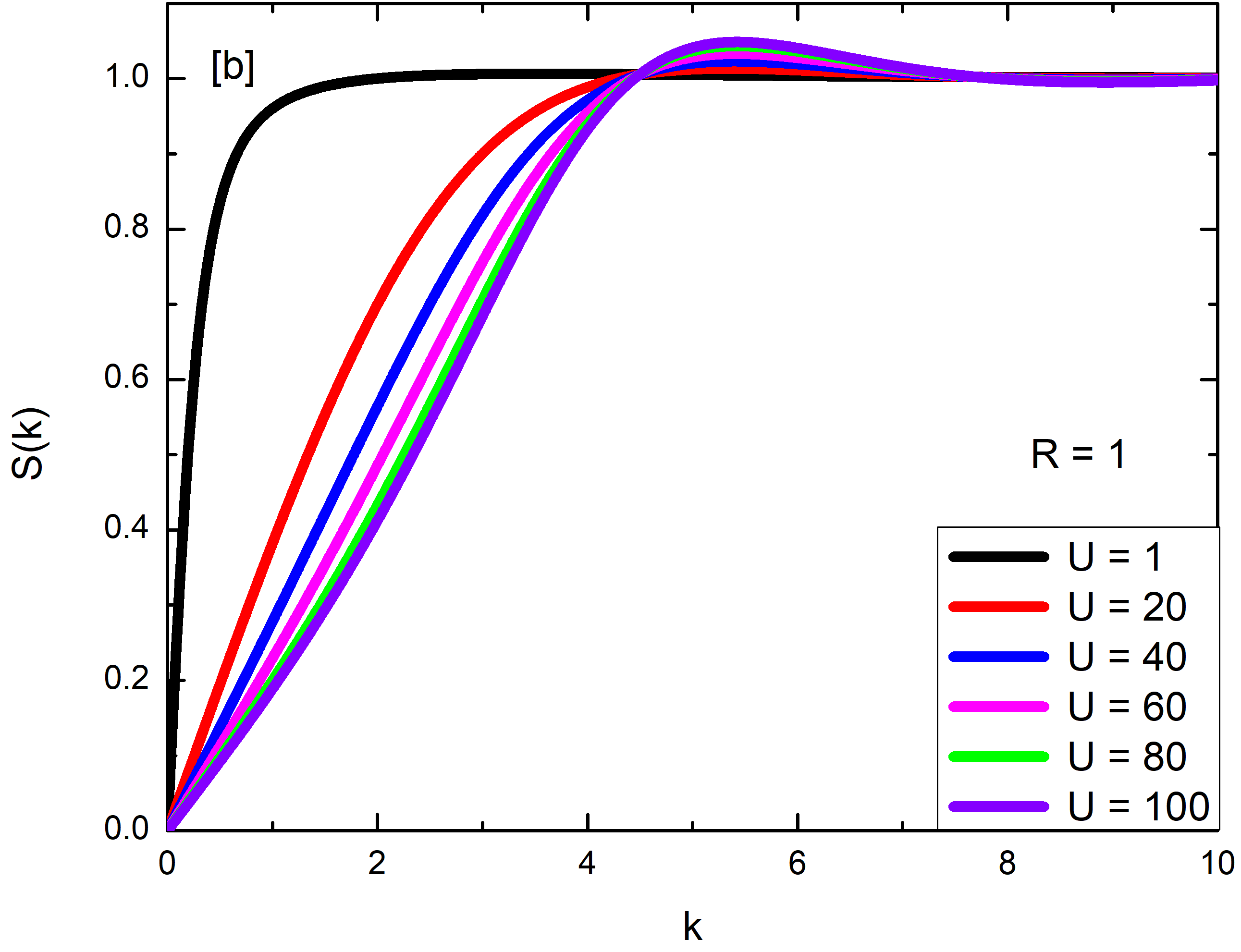}

\caption{
Collective excitation spectrum and corresponding static structure
factor for a quantum droplet in thermodynamic limit and
interaction range $R=1$. (a) Excitation spectrum $\omega(k)$ for
different nonlocal interaction strengths $U$. The spectrum remains
phonon dominated without a prominent roton minimum. The low-momentum
dispersion is linear and its slope increases with $U$, indicating an
enhancement of the phonon velocity, while at larger momenta the
dispersion approaches the free-particle limit. (b) Corresponding
static structure factor $S(k)$. A small peak appears at intermediate
momentum and grows slightly with increasing $U$, reflecting enhanced
density correlations, while $S(k)\to1$ at large momentum.
}
\label{fig:collective_R1}
\end{figure*}

where $V_0$ denotes the interaction strength, $R_c$ is the interaction
range, and $\Theta(x)$ is the Heaviside step function.

In our numerical calculations, we use dimensionless units to provide a general description of self-bound quantum droplets. The length, time, and energy are measured in units of $l_0$ (defined in Eq.~(9) of Ref.~\cite{petrov2015}), $ml_0^2/\hbar$, and $\hbar^2/(ml_0^2)$, respectively~\cite{code2}, where $m$ denotes the mass of the condensed atom. We reduce the coupled Gross--Pitaevskii (GP) equations for a binary Bose--Bose mixture to the effective single-component equation, Eq.~(\ref{single_gp}), following Ref.~\cite{petrov2015}. In this approach, the effective mean-field interaction strength is given by
\begin{equation}
\delta g=g_{12}+\sqrt{g_{11}g_{22}},
\end{equation}
where we consider equal intraspecies interaction strengths ($g_{11}=g_{22}$), an attractive interspecies interaction $g_{12}(<0)$, and equal populations in the two components, $N_1=N_2=N/2$, where $N_1$ and $N_2$ denote the particle numbers of the first and second species, respectively. The dimensionless extended Gross--Pitaevskii (GP) equation can then be written as~\cite{petrov2015,macri2013}

\begin{equation}
\begin{aligned}
i\frac{\partial \psi}{\partial t}
&=
\left[
-\frac{1}{2}\nabla^2
-3|\psi|^2
+\frac{5}{2}|\psi|^3
\right.\\
&\qquad\left.
+\int U(|\mathbf r-\mathbf r'|)
|\psi(\mathbf r',t)|^2 d\mathbf r'
\right]\psi ,
\end{aligned}
\label{single_gp}
\end{equation}

where the first term on the right-hand side represents the  kinetic energy, the second term corresponds to the effective attractive mean-field interaction with $\delta g=-3$, the third term is the repulsive Lee--Huang--Yang (LHY) correction, which is essential for stabilizing the quantum droplet by balancing the attractive mean-field interaction and preventing collapse, and the last term describes the finite-range interaction. The detailed derivation of the LHY term can be found in Ref.~\cite{LimaPelster2012}. The scaled step potential in equation (\ref{single_gp}) is

\begin{equation}
U(r)=U_0\,\Theta(R-r),
\end{equation}

where
\[
U_0=\frac{m l_0^2}{\hbar^2}V_0,\qquad
R=\frac{R_c}{l_0},
\]
are the dimensionless interaction strength and interaction range, respectively. The condensate wave function satisfies the normalization condition

\begin{equation}
\int |\psi(\mathbf r,t)|^2\, d\mathbf r = N,
\end{equation}

where $N$ is the total number of particles in the system.
\subsection{Numerical Method}

The extended Gross–Pitaevskii equation with nonlocal interaction is
solved numerically using the split-step Crank–Nicolson method
(SSCN). The nonlocal convolution term is computed in Fourier
space using the Fast Fourier Transform (FFT) implemented via
the FFTW library \cite{code1,code2,code3}.
The nonlocal interaction term has the convolution form

\begin{equation}
\Phi(\mathbf r)
=
\int U(|\mathbf r-\mathbf r'|)
|\psi(\mathbf r')|^2 d\mathbf r' .
\end{equation}

Using the convolution theorem, this expression can be evaluated efficiently
in momentum space as

\begin{equation}
\Phi(\mathbf r)
=
\mathcal{F}^{-1}
\left[
\tilde U(\mathbf k)
\tilde n(\mathbf k)
\right],
\end{equation}

where

\begin{equation}
\tilde n(\mathbf k)
=
\mathcal{F}
\left[
|\psi(\mathbf r)|^2
\right]
\end{equation}

is the Fourier transform of the density distribution.

For the step potential, the Fourier transform can be calculated analytically \cite{code1},

\begin{equation}
\tilde U(k)
=
4\pi V_0
\frac{\sin(kR)-kR\cos(kR)}{k^3}.
\label{FT_step}
\end{equation}

Forward and inverse Fourier transforms are computed numerically using the
FFTW library. During time propagation, the kinetic operator is evaluated in
momentum space while the nonlinear terms are evaluated in coordinate space.
The ground state is obtained by imaginary-time evolution with normalization
imposed after each iteration to preserve the particle number.

\subsection{Collective Excitations}
 In the large-droplet (thermodynamic) limit,  the collective excitation spectrum can be well described within the homogeneous Bogoliubov approximation applied to the bulk region of the droplet. This treatment is consistent with Ref.~\cite{PRA102053303}, where the surface excitation energy, $\omega_s$, scales as $N^{-1/2}$ and therefore becomes negligible for sufficiently large particle numbers ($N$). Accordingly, the bulk properties dominate the excitation spectrum in this limit. 
To investigate the stability and elementary excitations of the system \cite{collective},
we consider small perturbations around a stationary condensate state.
The time-independent solution of the scaled equation can be written as

\begin{equation}
\psi(\mathbf r,t)=\psi_0 e^{-i\mu t},
\end{equation}

where $\psi_0$ is the equilibrium condensate amplitude and $\mu$
is the chemical potential. The corresponding condensate density is
$n_0=|\psi_0|^2$.

Elementary excitations  are studied by introducing small fluctuations
around the stationary state ($\psi \to \psi+\delta\psi $) using the Bogoliubov ansatz \cite{avra_ryd, avra_surf}

\begin{equation}
\psi(\mathbf r,t)
=
\left[
\psi_0
+
u_{\mathbf k}e^{i(\mathbf k\cdot\mathbf r-\omega t)}
+
v_{\mathbf k}^*e^{-i(\mathbf k\cdot\mathbf r-\omega t)}
\right]e^{-i\mu t},
\label{excitation}
\end{equation}

where $u_{\mathbf k}$ and $v_{\mathbf k}$ denote the excitation
amplitudes and are assumed to be small compared with the condensate
amplitude $|\psi_0|$.

Substituting this expression into the GP equation (\ref{single_gp}) and retaining
only linear terms in the perturbations leads to the Bogoliubov–de Gennes
equations for the excitation amplitudes $u_k$ and $v_k$,

\begin{equation}
\omega u_k
=
(\epsilon_k+A_k)u_k + B_k v_k ,
\end{equation}

\begin{equation}
-\omega v_k
=
(\epsilon_k+A_k)v_k + B_k u_k ,
\end{equation}

which describe the coupling between particle- and hole-like
excitations in the condensate.

The kinetic contribution to the excitation energy is given by the
 dispersion of the free-particles

\begin{equation}
\epsilon_k=\frac{k^2}{2}.
\end{equation}

The coefficients $A_k$ and $B_k$ arise from the linearization of the
interaction terms in the scaled Gross–Pitaevskii equation and encode
the effects of mean-field interactions, quantum fluctuations, and
finite-range interactions. Their explicit form is

\begin{equation}
A_k=B_k
=
-3n_0 + n_0\tilde V(k) + \frac{15}{8}n_0^{3/2}.
\end{equation}

Here $\tilde V(k)$ denotes the Fourier transform of the soft-core
interaction potential, which introduces a momentum-dependent
interaction contribution to the excitation spectrum.

Diagonalization of the Bogoliubov system yields the dispersion relation
of the collective excitations  (Similar treatments for different systems can be found in Refs.~\cite{exe,moniri,avra_PTB}.)

\begin{equation}
\omega(k)
=
\sqrt{
\frac{k^2}{2}
\left[
\frac{k^2}{2}
+
2n_0(-3+\tilde V(k))
+
\frac{15}{4}n_0^{3/2}
\right]
}.
\end{equation}

This spectrum determines the stability of the condensate and
reveals the possible emergence of phonon, maxon, and roton features
depending on the strength and range of the interaction.
\subsection{Static Structure Factor}

The static structure factor characterizes density correlations
and can be obtained within the Bogoliubov theory from density fluctuations \cite{st},
\begin{figure*}
\centering
\includegraphics[width=0.45\textwidth]{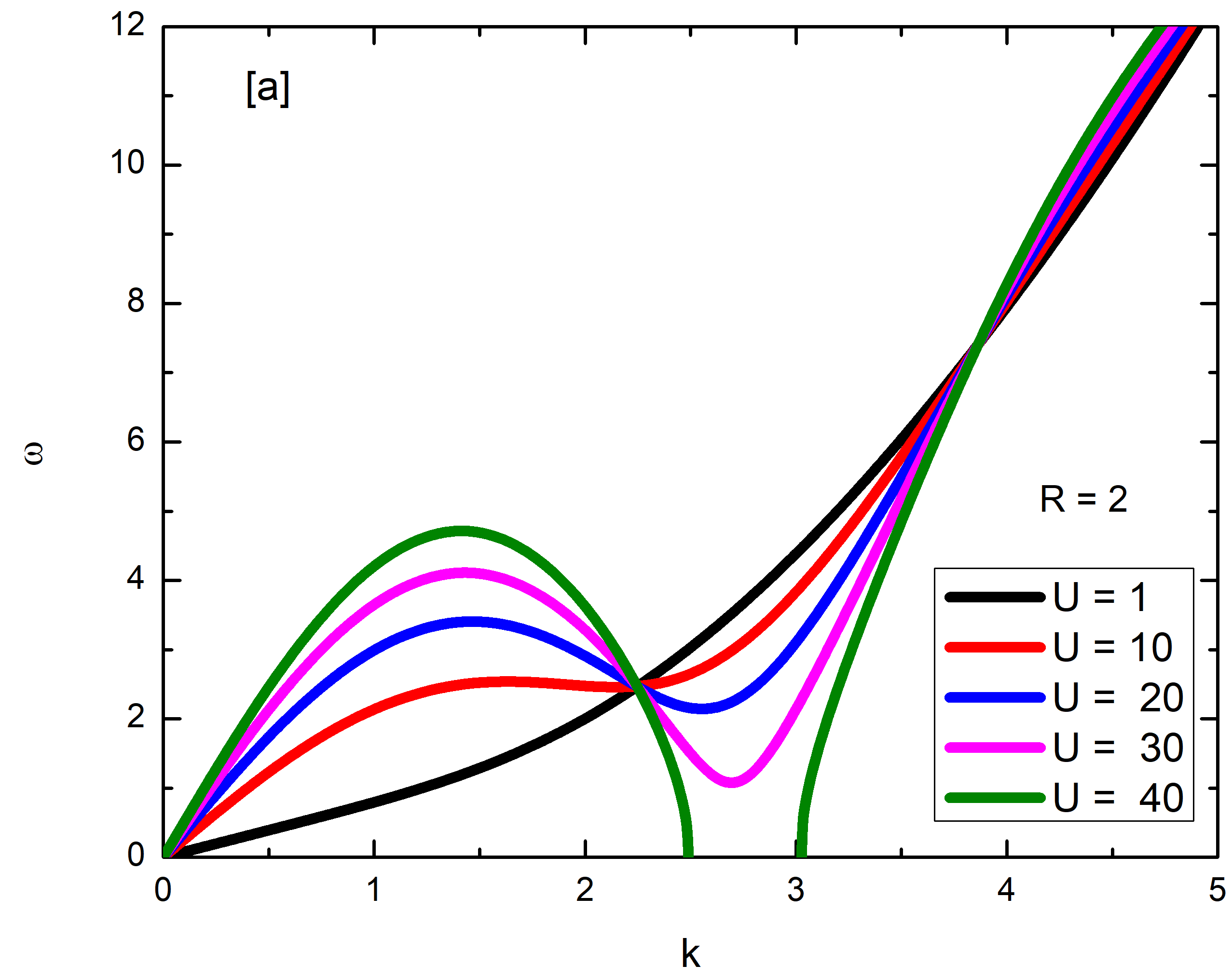}
\includegraphics[width=0.45\textwidth]{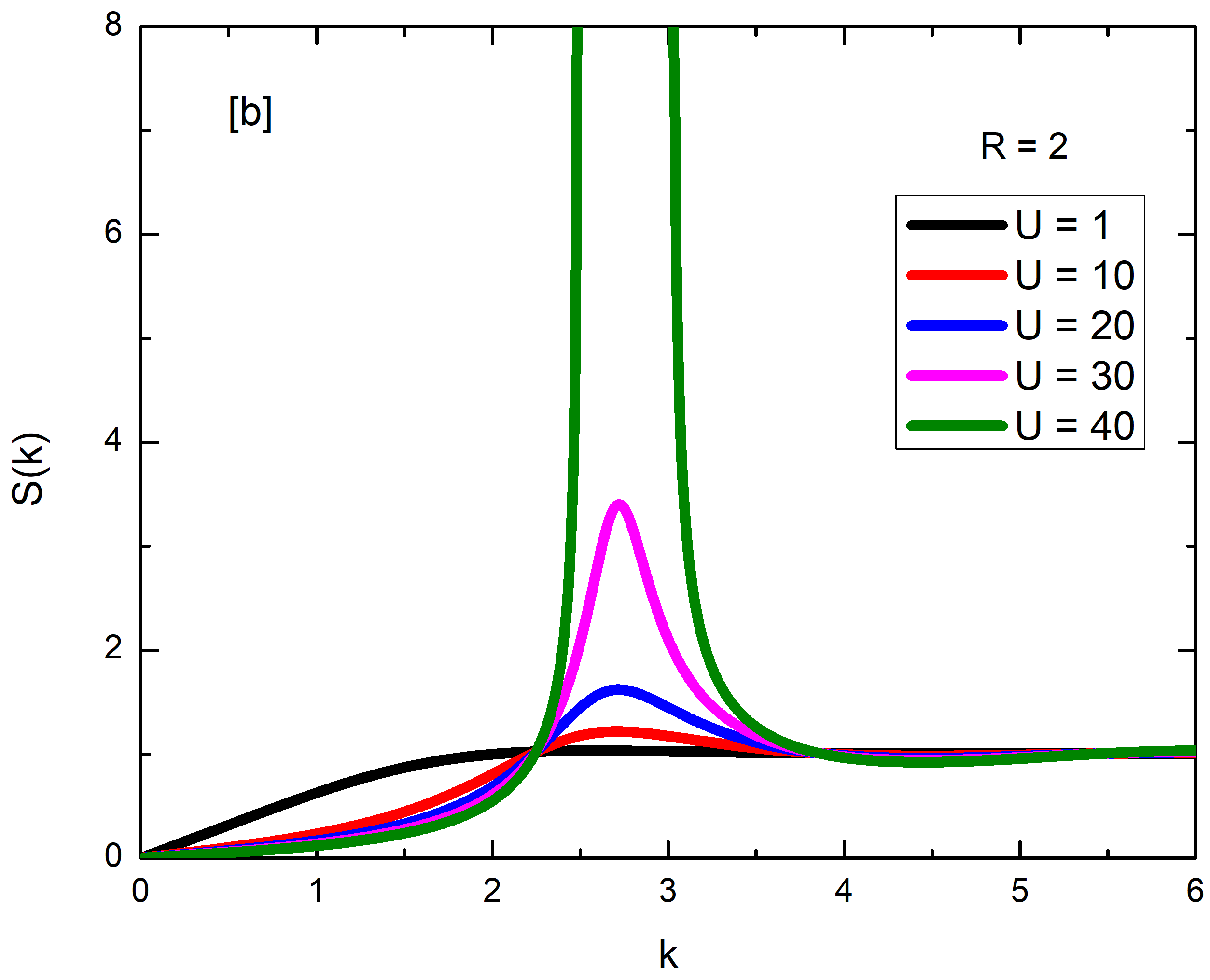}

\caption{Collective excitation spectrum and corresponding static structure factor for interaction range $R=2$  in thermodynamic limit . (a) Excitation spectrum $\omega(k)$ for different nonlocal interaction
strengths $U$. For weak interaction ($U=1$) the spectrum shows linear
phononic dispersion at small momentum. With increasing $U$, a roton-like
minimum develops at finite momentum and progressively softens,
indicating the approach to roton instability \cite{moniri,roton_ins}. The slope of the phonon
branch also increases, reflecting an enhancement of the phonon velocity. (b) Corresponding static structure factor $S(k)$. A peak emerges and
grows at finite momentum with increasing $U$, signaling enhanced
density correlations associated with the roton mode, while $S(k)\to1$
at large momentum.}
\label{fig:roton_excitation}
\end{figure*}
\begin{equation}
S(k)=\frac{1}{N}\langle \delta n_{\mathbf k}\delta n_{-\mathbf k}\rangle .
\end{equation}

Using the Bogoliubov expansion equation (\ref{excitation}) the density fluctuation becomes
\begin{equation}
\delta n_k=\sqrt{n_0}(u_k+v_k).
\end{equation}

Employing the Bogoliubov normalization, $u_k^2-v_k^2=1$, leads to

\begin{equation}
S(k)=(u_k+v_k)^2.
\end{equation}

Using the Bogoliubov solution, this reduces to the Feynman relation

\begin{equation}
S(k)=\frac{\epsilon_k}{\omega(k)},
\end{equation}

where $\epsilon_k=k^2/2$ in the scaled units and
$\omega(k)$ is the excitation spectrum derived previously.

\section{Results and Discussion} 

The density profiles shown in Fig.~\ref{fig:droplet_density} illustrate how the presence of a finite-range interaction modifies the internal structure of the droplet. In the absence of the nonlocal interaction (Fig.~1a), the droplet exhibits a nearly uniform bulk density with a well-defined boundary, characteristic of a self-bound state stabilized by the balance between mean-field attraction and LHY repulsion. When the nonlocal interaction is introduced, the density distribution begins to develop spatial correlations. For moderate interaction strength, the droplet expands slightly while maintaining a nearly uniform interior density (Fig.~1b).  The density decreases rapidly but continuously near the droplet surface, giving the appearance of an abrupt edge in the plotted scale.  As the interaction range increases, density modulations gradually emerge due to the increasing influence of the finite-range interaction (Fig.~1c). The competition between the contact and finite-range interactions favors a spatially modulated density profile. The collective excitation spectrum of a quantum droplet with  Pöschl--Teller interaction was studied in Ref.~\cite{avra_PTD}. In the present work, we consider a step-function interaction whose Fourier transform is given in Eq.~(\ref{FT_step}). Since the two interaction potentials exhibit different momentum-space behavior, their contributions to the excitation spectrum and density profile are expected to be different.  At sufficiently strong interaction, the density profile becomes highly irregular and fragmented, suggesting that the uniform droplet configuration is destabilized by the finite-range interaction (Fig.~1d). 

To understand the origin of these spatial modulations we analyze the collective excitation spectrum obtained from Bogoliubov theory (Figs.~2 and 3) in the thermodynamic limit. For a small interaction range ($R=1$), the excitation spectrum remains dominated by phonon-like modes at low momentum. The linear dispersion in the long-wavelength limit reflects the compressible nature of the quantum droplet, where the slope of the dispersion determines the sound velocity. Increasing the interaction strength leads to a steeper linear branch, indicating an enhancement of the phonon velocity due to stronger effective interactions (Fig.~2a). The static structure factor provides complementary information about density correlations. For weak interactions, the structure factor remains smooth and approaches unity at large momentum, consistent with the free-particle limit. As the interaction strength increases, a small peak begins to develop at intermediate momentum, indicating enhanced density correlations induced by the nonlocal interaction (Fig.~2b). A qualitatively different behavior emerges when the interaction range is increased. In this regime, the excitation spectrum develops a roton-like minimum at finite momentum. As the interaction strength increases, the roton energy progressively decreases and eventually approaches zero, signaling the onset of a dynamical instability \cite{moniri, roton_ins,dynamic}. This softening of the roton mode indicates that the system becomes susceptible to density modulations with a characteristic wavelength determined by the roton momentum (Fig.~3a). The behavior of the static structure factor is consistent with this interpretation. As the roton minimum develops, a pronounced peak appears in the structure factor at the corresponding momentum, reflecting strong density correlations in the droplet. The growth of this peak provides a clear signature of the increasing tendency of the system toward spatial ordering (Fig.~3b).

\section{Conclusion} In summary, we have investigated the emergence of roton instability in self-bound quantum droplets interacting via a finite-range soft-core potential. Using an extended Gross–Pitaevskii equation including Lee–Huang–Yang corrections and a nonlocal interaction term, we analyze both the ground-state density distribution and the collective excitation spectrum. Our results show that finite-range interactions significantly modify the excitation spectrum of the droplet and can lead to the formation of a roton minimum at finite momentum. As the interaction strength or interaction range increases, the roton energy softens and eventually approaches zero, indicating the onset of a dynamical instability associated with density modulation. The development of the roton mode is accompanied by pronounced peaks in the static structure factor, demonstrating the growth of density correlations in the system. These findings highlight the crucial role of finite-range interactions in determining the stability and collective behavior of quantum droplets. The present model therefore provides a minimal framework for exploring roton physics in self-bound Bose systems and may serve as a useful reference for future studies of Rydberg-dressed and other soft-core interacting quantum gases.

\end{document}